\documentclass[twocolumn,pra,showpacs,amsmath,amssymb,amsfonts,superscriptaddress]{revtex4}

\usepackage{graphicx}

\makeatletter
\newcommand{\instopt}{Laboratoire Charles Fabry de l'Institut d'Optique, CNRS
et Universit\'{e} Paris Sud 11 \\ Campus Polytechnique, RD 128,
91127 Palaiseau, France}
\newcommand{\syrte}{SYRTE, Observatoire de Paris, CNRS, UPMC \\ 61 avenue de l'Observatoire,
75014 Paris, France}

\newcommand{\psil}{\psi_{\ell}}

\newcommand{\omp}{\omega_{\perp}}

\makeatother
\begin{document}

\title{Theoretical tools for atom laser beam propagation}

\author{J.-F. Riou\footnote{Present address: Physics Department of Penn State University, 104 Davey Laboratory, Mailbox 002, University Park, PA 16802, U.S.A.}}
\email{jfriou@phys.psu.edu}
\affiliation{\instopt}
\author{Y. Le Coq}
\affiliation{\syrte}
\author{F. Impens}
\affiliation{\syrte}
\author{W. Guerin\footnote{Present address: Institut Non Lin\'{e}aire de Nice, 1361 route des Lucioles, 06560 Valbonne, France}}
\affiliation{\instopt}
\author{C. J. Bord\'{e}}
\affiliation{\syrte}
\author{A. Aspect}
\affiliation{\instopt}
\author{P. Bouyer}
\affiliation{\instopt}

\date{\today}

\begin{abstract}
We present a theoretical model for the propagation of
non self-interacting atom laser beams. We start from a general propagation
integral equation, and we use the same approximations as in photon
optics to derive tools to calculate the atom laser beam propagation.
We discuss the approximations that allow to reduce the general
equation whether to a Fresnel-Kirchhoff integral calculated 
by using the stationary phase method, or to the eikonal. 
Within the paraxial approximation, we also introduce
the $ABCD$ matrices formalism and the beam quality factor. As an example, we
apply these tools to analyse the recent experiment by Riou et al. [Phys. Rev. Lett. ${\bf 96}$, 070404 (2006)].
\end{abstract}


\date{\today}
\maketitle

\section*{Introduction}
\label{Intro}

Matter-wave optics, where a beam of neutral atoms is considered for
its wave-like behavior, is a domain of considerable studies, with
many applications, ranging from atom lithography to atomic clocks
and atom interferometer \footnote{See for example App. Phys. B
\textbf{84}(4), special issue \textit{Quantum Mechanics for Space
Application: From Quantum Optics to Atom Optics and General
Relativity} (2006).}. The experimental realization of coherent
matter-wave - so called atom lasers
\cite{Mewes1997,Anderson1998,Bloch1999,Hagley1999,Cennini2003,Robins2006,Guerin2006}
- which followed the observation of Bose-Einstein condensation
put a new perspective to
the field by providing the atomic analogue to photonic laser beams.

Performant theoretical tools for characterizing the propagation
properties of matter waves and their manipulation by atom-optics
elements are of prime interest for high accuracy applications, as
soon as one needs to go beyond the proof-of-principle experiment.
In the scope of partially coherent atom interferometry, and for
relatively simple (\textit{i.e.} homogenous) external potentials,
many theoretical works have been developed
\cite{Borde2001,Storey1994,Antoine2003,Bongs2006} and applied
successfully \cite{Oates2006,Wilpers2007}. All these tools
essentially address the propagation of an atomic wavepacket. For
fully coherent atom-laser beams, most theoretical investigations
focused on  the dynamics of the outcoupling
\cite{Ballagh1997,Naraschewski1997,Steck1998,Zhang1998,Kneer1998,Schneider1999,Band1999,Edwards1999,Graham1999,Schneider2000,Robins2001,Gerbier2001,Robins2004,Robins2005}
and the quantum statistical properties of the output beam
\cite{Moy1997,Moy1999,Jack1999,Bradley2003,
Japha1999,Choi2000,Japha2002,Ruostekoski2003}. Some works
specifically addressed  the spatial shape of the atom laser beam
\cite{Busch2002a,Kramer2006}, but rely
essentially on numerical simulations or neglect the influence of
dimensionality and potential inhomogeneity. For realistic
experimental conditions, the 3D external potential is
inhomogeneous, and full numerical simulation become particularly
cumbersome. One thus needs a simplified analytical theoretical
framework to handle the beam propagation.

Following our previous work  \cite{LeCoq2001,Riou2006}, we present
here in detail a simple but general framework for the propagation
of atom laser beams in inhomogeneous media. We show how several
theoretical tools from classical optics can be adapted for
coherent atom-optics. We address three major formalisms used in
optics : the eikonal approximation, the Fresnel-Kirchhoff integral,
and the $ABCD$ matrices formalism in the paraxial approximation.

The first part of the paper gives an overview of these theoretical
tools for atom laser beam propagation. In the first section, we
introduce the integral equation of the propagation and its
time-independent version. We present in the second section
different ways of dealing with the time-independent propagation of
the matter wave. First, the time-independent propagator is computed
using the stationary phase approximation. Then, we show that two
approximations -the eikonal and the paraxial approximation-, which
apply in different physical contexts, can provide a more tractable
treatment than the general integral equation.
In the second part, we show in practice how to use these methods
in the experimental case of \cite{Riou2006} with a rubidium
radiofrequency-coupled atom laser. Some of these methods have
recently been used also for a metastable helium atom laser
\cite{Dall2007} as well as for a Raman-coupled atom laser
\cite{Jeppesen2007}.

\section{Analytical propagation methods for matter waves}
\label{Generalframework}

\subsection{Matter wave weakly outcoupled from a source}

\subsubsection{Propagation equation}

We consider a matter wave $\psi_{\ell}({\bf r},t)$ outcoupled from a source $\psi_{\mathrm{s}}({\bf r},t)$. We note $V_i({\bf r},t)$ ($i=\{\ell,\mathrm{s}\}$), the external potential in which each of them evolves. We also introduce a coupling term  $W_{ij}({\bf r},t)$ between $\psi_i$ and $\psi_j$. In the mean-field approximation, such system is described by a set of two coupled Gross-Pitaevskii equations, which reads
\begin{equation}
i\hbar\,\partial_t\psi_i\!=\!\!\left[-\frac{\hbar^2}{2m}\Delta+V_i+\sum_{k=\ell,\mathrm{s}} g_{ki}\left|\psi_k\right|^2\right]\!\! \psi_i +  W_{ij}\psi_j\,.
\label{eq:sysgeneral}
\end{equation}
In this equation, $g_{ik}$ is the mean-field interaction strength between states $i$ and $k$. The solution of such equations is not straightforward, mainly due to the presence of a nonlinear
mean-field term. However, in the case of propagation of
matter waves which are weakly outcoupled from a source, one can
greatly simplify the treatment \cite{Gerbier2001}. Indeed, the
weak-coupling assumption implies the two following points:
\begin{itemize}
\item The evolution of the source wave-function is unaffected by
the outcoupler, \item the extracted matter wave is sufficiently
diluted to make self-interactions negligible.
\end{itemize}

The former differential system can then be rewritten as:
\begin{eqnarray}
i\hbar\,\partial_t\psi_{\mathrm{s}}\!=\!\!\left[-\frac{\hbar^2}{2m}\Delta+V_{\mathrm{s}}+
g_{\mathrm{s}\mathrm{s}}\left|\psi_{\mathrm{s}}\right|^2\right]
\psi_{\mathrm{s}}\,, \label{eq:sourcegeneral} \\
i\hbar\,\partial_t\psi_{\ell}\!=\!\!\left[-\frac{\hbar^2}{2m}\Delta+V_{\ell}+
g_{\mathrm{s}{\ell}}\left|\psi_\mathrm{s}\right|^2\right]\!\!
\psi_{\ell} +  W_{{\ell}\mathrm{s}}\psi_{\mathrm{s}}\,.
\label{eq:outcouplgeneral}
\end{eqnarray}

The source wave-function $\psi_{\mathrm{s}}({\bf r},t)$ now obeys a single differential equation \eqref{eq:sourcegeneral}, and can thus be determined independently.  The remaining nonlinear term $|\psi_\mathrm{s}|^2$ in Eq. \eqref{eq:outcouplgeneral}, acts then as an external potential for the propagation of $\psi_{\ell}$. This last equation is thus a Schr\"{o}dinger equation describing the evolution of the outcoupled matter wave in the total potential $V({\bf r},t)$ in presence of a source term $\rho({\bf r},t)$,
\begin{equation}
    i\hbar\partial_t\psi_{\ell} = H_{\mathbf{r}} \psi_{\ell} + \rho\,,
    \label{eq:Schrodinger}
\end{equation}
where
\begin{subequations}
\begin{align}
H_{\mathbf{r}}&=-\frac{\hbar^2}{2m}\Delta_{\mathbf{r}}+V\,,\label{eq:defSchrodingera}\\
V&=V_{\ell}+ g_{\mathrm{s}{\ell}}\left|\psi_\mathrm{s}\right|^2\,,\label{eq:defSchrodingerb}\\
\rho&=W_{{\ell}\mathrm{s}}\psi_{\mathrm{s}}\label{eq:defSchrodingerc}\,.
\end{align}
\end{subequations}

\subsubsection{Integral equation}

The evolution between times $t_0$ and $t$ ($t>t_0$) of the solution $\psi_{\ell}$ of Eq. \eqref{eq:Schrodinger} in a given volume $\mathcal{V}$ delimited by a surface $\mathcal{S}$, is expressed by an implicit integral \cite{Barton1989}
\begin{multline}
    \psi_{\ell}(\mathbf{r},t) =  \int_{\mathcal{V}}d\mathbf{r'}\mathcal{G}(\mathbf{r},\mathbf{r'},t-t_0)\,\psi_{\ell}(\mathbf{r'},t_0)\\
    + \frac{i\hbar}{2m} \int^{t}_{t_0}dt' \int_{\mathcal{S}}d\mathbf{S'}\cdot\left[\mathcal{G}(\mathbf{r},\mathbf{r'},t-t')\nabla_{\mathbf{r'}}\psi_{\ell}(\mathbf{r'},t')\right.\\
    -\psi_{\ell}(\mathbf{r'},t')\left.\nabla_{\mathbf{r'}}\mathcal{G}(\mathbf{r},\mathbf{r'},t-t')\right]\\
            + \frac{1}{i\hbar} \int^{t}_{t_0}dt'                            \int_{\mathcal{V}}d\mathbf{r'}\mathcal{G}(\mathbf{r},\mathbf{r'},t-t')\,\rho(\mathbf{r'},t')\,,
        \label{eq:MagicRule}
\end{multline}
where $d\mathbf{S'}$ is the outward-oriented elementary normal vector to the surface $\mathcal{S}$. We have introduced the time-dependent Green function $\mathcal{G}(\mathbf{r},\mathbf{r'},\tau)$ which verifies
\begin{equation}
\left[i\hbar\partial_{\tau}-H_{\mathbf{r}}\right]\mathcal{G}=i\,\hbar\,\delta(\tau)\,\delta(\mathbf{r}-\mathbf{r'})\,,
\end{equation}
and is related to the propagator $\mathcal{K}$ of the Schr\"{o}dinger equation via a Heaviside function $\Theta$ ensuring causality,
\begin{equation}
\mathcal{G}(\mathbf{r},\mathbf{r'},\tau)=\mathcal{K}(\mathbf{r},\mathbf{r'},\tau)\,\Theta(\tau)\,.
\end{equation}

Eq. \eqref{eq:MagicRule} states that, after the evolution time
$t-t_0$, the value of the wave function is the sum of three terms,
the physical interpretation of which is straightforward. The first one
corresponds to the propagation of the initial condition
$\psi_{\ell}(\mathbf{r'},t_0)$ given at any position in the volume
$\mathcal{V}$. The second one takes into account the propagation
of the wave function taken at the surrounding surface
$\mathcal{S}$, and is non-zero only if $\mathcal{V}$ is finite.
This term takes into account any field which enters or leaks out
of $\mathcal{V}$. Finally, the last term expresses the
contribution from the source.

Eq. \eqref{eq:MagicRule} can be successfully applied to describe
the propagation of wavepackets in an atom interferometer as
described in \cite{Borde1990}. Nevertheless, the propagation of a
continuous atom laser, the energy of which is well defined, can be
described with a time-independent version of Eq.
\eqref{eq:MagicRule}, that we derive below.

\subsubsection{Time-independent case}

We consider a time-independent hamiltonian $H_{\mathbf{r}}$ and a stationnary source
\begin{equation}
\rho(\mathbf{r},t) = \rho(\mathbf{r}) \exp{(-iEt/\hbar)}\,.
\label{eq:timeindeprho}
\end{equation}
We thus look for stationnary solutions of Eq.
\eqref{eq:Schrodinger} with a given energy $E$,
\begin{equation}
\psi_{\ell}(\mathbf{r},t) = \psi_{\ell}(\mathbf{r}) \exp{(-iEt/\hbar)}\,.
\label{eq:timeindeppsi}
\end{equation}
When $t_0 \rightarrow -\infty$, Eq. \eqref{eq:MagicRule} then
becomes time-independent:
\begin{multline}
    \psi_{\ell}(\mathbf{r}) =                          \frac{1}{i\hbar}\int_{\mathcal{V}}d\mathbf{r'}\mathcal{G}_E(\mathbf{r},\mathbf{r'})\rho(\mathbf{r'})\\
+\!\frac{i\hbar}{2m}\!\int_{\mathcal{S}}\!d\mathbf{S'}\!\cdot\!\left[\mathcal{G}_E(\mathbf{r},\mathbf{r'})\nabla_{\mathbf{r'}}\psi_{\ell}(\mathbf{r'})-\psi_{\ell}(\mathbf{r'})\nabla_{\mathbf{r'}}\mathcal{G}_E(\mathbf{r},\mathbf{r'})\right]\!,
    \label{eq:MagicRuleIndepT}
\end{multline}
where $\mathcal{G}_E$ is the time-independent propagator related to $\mathcal{K}$ via
\begin{equation}
    \mathcal{G}_E(\mathbf{r},\mathbf{r'}) = \int_{0}^{+\infty} d\tau\,\mathcal{K}(\mathbf{r},\mathbf{r'},\tau)\, e^{i\frac{E\tau}{\hbar}}\,.
    \label{eq:IntegralDefinitionGE}
\end{equation}
Note that the first term of Eq. \eqref{eq:MagicRule} vanishes in
the time-independent version of the propagation integral
equation as $\mathcal{K}(\mathbf{r},\mathbf{r'},\tau)\rightarrow
0$ when $\tau\rightarrow\infty$. The second term of Eq.
\eqref{eq:MagicRuleIndepT} is the equivalent for matter waves of
what is known in optics as the Fresnel-Kirchhoff integral
\cite{Born2002}.

\subsection {Major approximations for atom laser beam propagation}

\subsubsection{Independent treatment of a succession of potentials}

As an optical wave can enter different media (free space, lenses...) separated by surfaces, matter waves can propagate in different parts of space, where they experience potentials of different nature.
For instance, when one considers an atom laser outcoupled from a condensate as in the example of part II, the beam initially interacts with the Bose-condensed atoms and abruptly propagates in free space outside of the condensate.
The expression of the propagator in whole space would then be needed to
use the equation \eqref{eq:MagicRuleIndepT}. Most generally,
such calculation requires to apply the Feynmann's path integral
method, either numerically or analytically \cite{Feynman1965}. For
example, the time-dependent propagator $\mathcal{K}$ can be
analytically expressed in the case of a continuous potential which
is at most quadratic, by using the Van Vleck's formula
\cite{vanVleck1928},
or the ABCD formalism \cite{Borde1990}. However such expressions
fail to give the global propagator value for a piecewise-defined quadratic potential.

As in classical optics, we can separate the total evolution of a
monochromatic wave in steps, each one corresponding to one
homogeneous potential. This  step-by-step approach stays valid as
long as one can neglect any reflection on the interface between
these regions as well as feedback from one region to a previous
one. In this approach,  each interface is considered as a surface
source term for the propagation in the following media. It allows
us to calculate $\mathcal{K}$ explicitly in every part of space as
long as the potential in each region remains at most quadratic, which
we will assume throughout this paper.

\subsubsection{The time-independent propagator in the stationary phase approximation}
\label{GreenGeneral}

Whereas the expression of $\mathcal{G}_E$ is well known for free
space and linear potentials \cite{Borde2001,Kramer2006}, to our
knowledge, there is no analytical expression for the inverted
harmonic potential, which plays a predominant role in an atom
laser interacting with its source-condensate. We thus give in
the following a method to calculate the time-independent
propagator $\mathcal{G}_E$ in any up to quadratic potential.

Since $\mathcal{K}$ is analytically known in such potentials, we use the definition of
$\mathcal{G}_E$ as its Fourier transform (Eq.
\ref{eq:IntegralDefinitionGE}). The remaining integral over time
$\tau$ is calculated via a stationary phase method
\cite{Born2002}, taking advantage that $\mathcal{K}$ is a rapidly
oscillating function. We write the time-dependent propagator as
\begin{equation}
\mathcal{K}(\mathbf{r},\mathbf{r'}, \tau) = \mathcal{A}(\tau) \exp\left[i\phi(\mathbf{r},\mathbf{r'}, \tau)\right]\,.
\label{eqKAPhi}
\end{equation}
We introduce $\tau_n$ as the positive real solution(s) of
\begin{equation}
\partial_\tau \phi(\mathbf{r},\mathbf{r'},\tau_n)=-E/\hbar\,,
\label{eq:defstat1}
\end{equation}
which correspond(s) to the time(s) spent on classical path(s) of energy $E$ connecting $\mathbf{r'}$ to $\mathbf{r}$.
We develop $\phi$ to the second order around $\tau_n$,
\begin{equation}
\phi(\tau)\simeq \phi(\tau_n)+\frac{\partial \phi}{\partial \tau}\Bigr|_{\tau_n}\!\!\!(\tau-\tau_n)+\frac{\partial^2 \phi}{\partial \tau^2}\Bigr|_{\tau_n}\!\!\!\frac{(\tau-\tau_n)^2}{2}\,.
\label{eq:devphasestat1}
\end{equation}
Using the last development in the integral \eqref{eq:IntegralDefinitionGE}, and assuming that the enveloppe $\mathcal{A}(\tau)$ varies smoothly around $\tau_n$, we can express $\mathcal{G}_E$ as
\begin{equation}
\mathcal{G}_E^{(1)}  \simeq  \sum_n \sqrt{\frac{2i\pi}{\phi''(\tau_n)}} \, \mathcal{K}(\tau_n)\, \exp\left(i\frac{E\tau_n}{\hbar}\right)\,.
\label{eq:GE1}
\end{equation}
Such approach is valid as long as stationary points $\tau_n$ exist
and their contribution can be considered independently: Eq. \ref{eq:GE1} fails if the stationary points are
too close to each other. We can estimate the validity of our
approach by defining an interval
$\mathcal{I}_n=[\tau_n-\theta_n;\tau_n+\theta_n]$ in which the
development around $\tau_n$ contributes to more than $\beta=90\%$
to the restricted integral. For $\theta$ large enough, we can use
\cite{Brillouin1916},
\begin{equation}
\left|\int_{-\theta}^{\theta}dx\,\exp\left[i z \frac{x^2}{2}\right]-\sqrt{\frac{2i\pi}{z}}\,\right| \sim \frac{2}{\left|z\,\theta\right|},
\end{equation}
and obtain $\theta_n$
\begin{equation}
\theta_n=\frac{1}{1-\beta}\sqrt{\frac{2}{\pi \phi''(\tau_n)}}\,.
\end{equation}
The validity condition is thus
$|\tau_{n}-\tau_{n+1}|\geq\theta_{n}+\theta_{n+1}$.

If \eqref{eq:GE1} is not valid, a better approximation consists
then in developing $\phi$ to higher order around a point which is
inbetween successive $\tau_n$. The simplest choice is to take the one which cancels
$\phi''$, and to choose stationnary points $\tau_k$ which verify
\begin{equation}
\partial_\tau^2 \phi(\mathbf{r},\mathbf{r'},\tau_k)=0\,.
\label{eq:defstat2}
\end{equation}
We thus develop $\phi$ to the third order around $\tau_k$,
which leads to the following expression of $\mathcal{G}_E$,
\begin{equation}
\mathcal{G}_E^{(2)}  \simeq \frac{2\pi
\mathcal{K}(\tau_k)\, \exp\left(i\frac{E\tau_k}{\hbar}\right)}{\sqrt[3]{-\phi^{(3)}(\tau_k)/2}} \mathrm{Ai}\left( \frac{\phi'(\tau_k)+E/\hbar}{\sqrt[3]{\phi^{(3)}(\tau_k)/2}}\right) \,,
\label{eq:GE2}
\end{equation}
where Ai is the Airy function of the first kind \cite{Abramowitz1972}.

In practice, combining the use of $\mathcal{G}_E^{(2)}$ and $\mathcal{G}_E^{(1)}$ depending on the values of $\mathbf{r'}$ and $\mathbf{r}$ gives a good estimate of the time-independent propagator, as we will see in part II.

Although the above approach is quite general, further approximations can be made. In the region where diffraction can be neglected, one can describe the propagation with the eikonal approximation. When the propagation is in the paraxial regime, it is more appropriate to describe it with the paraxial $ABCD$ matrices, instead of using the general Kirchhoff integral.

\subsubsection{Eikonal propagation}
\label{sec:eikonal}

The purpose of this method, equivalent to the WKB approximation, is to give a semi-classical description of the propagation from a matter wave, given its value on a surface. Let us consider that we know the value of the wave function of energy $E$ on the surface $\mathcal{S}'$. To calculate its value on any other surface $\mathcal{S}$, the eikonal considers classical paths connecting $\mathcal{S}$ and $\mathcal{S}'$.
Let us write the wave function as
\begin{equation} \psi_{\ell}(\mathbf{r})=A(\mathbf{r})\exp\left[\,iS(\mathbf{r})/\hbar\,\right]\,.
\label{eq:psieikonal}
\end{equation}
The Schr\"{o}dinger equation on $\psi_{\ell}$ reduces to \cite{Messiah2000}
\begin{equation}
    \left\{
    \begin{aligned}
        \left|{\bf{\nabla_\mathbf{r}}} S\right|&=\frac{\hbar}{\lambdabar}\,,\\      {\nabla}_\mathbf{r}\!\cdot\!\left(A^2{\bf{\nabla}}S\right)&=0\,,\\
    \end{aligned}
    \right.
\label{eq:eikonal}
\end{equation}
where we have introduced the de Broglie wavelength
\begin{equation}
\lambdabar(\mathbf{r})=\frac{\hbar}{\sqrt{2m\left(E-V(\mathbf{r})\right)}}\,.
\end{equation}
The first equation is known in geometric optics as the eikonal equation \cite{Landau1975,Born2002}. The calculation consists in integrating the phase along the classical ray of energy $E$ connecting $\mathbf{r'}$ to $\mathbf{r}$, to obtain the phase on $\mathbf{r}$,
\begin{equation}
    S(\mathbf{r}) = \int_\mathbf{r'}^\mathbf{r}\!\!d\mathbf{u}\,\,\frac{\hbar}{\lambdabar(\mathbf{u})}+S(\mathbf{r'})\,.
 \label{eq:Seikonal}
\end{equation}

The second equation of system \eqref{eq:eikonal} corresponds to the conservation of probability density flux, and is equivalent to the Poynting's law in optics. Again, after integration along the classical path
connecting $\mathbf{r'}$ to $\mathbf{r}$, one obtains the amplitude on $\mathcal{S}$
\begin{equation}
A(\mathbf{r})=A(\mathbf{r'})\,\exp{\left(-\int_\mathbf{r'}^\mathbf{r}\!\!d\mathbf{u}\,\frac{\Delta S(\mathbf{u}) \lambdabar(\mathbf{u})}{2\hbar}\right)}\,.
 \label{eq:Aeikonal}
\end{equation}
Note that interference effects are included in this formalism: if several classical paths connect $\mathbf{r'}$ to
$\mathbf{r}$, their respective contributions add coherently to each other. Also, if some focussing points
exist, dephasings equivalent to the Gouy phase in optics appear and can be calculated following \cite{Born2002}.

Such semiclassical treatment is valid as long as one does not look
for the wave function value close to classical turning points, and
as long as transverse diffraction is negligible : transverse
structures of size $\Delta x$ must be large enough not to diffract
significantly, \textit{i.e.} $\Delta x^4 \gg (\hbar t/2m)^2$. This condition
restricts the use of the eikonal to specific regions of space
where the matter wave does not spend a too long time $t$. For
instance, this is the case for the propagation in the small region
of overlap with the BEC.

The eikonal can thus be used to deal with the first term of Eq.
\eqref{eq:MagicRuleIndepT}, and is equivalent to the development
of this integral around classical trajectories \footnote{The
stationary phase method could also provide a way to calculate this
integral, see \cite{PhDThesisRiou}.}.

\subsubsection{Paraxial propagation}\label{propagparax}

The paraxial regime applies as soon as the transverse wavevector
becomes negligible compared to the axial one.
It is for instance the case after some propagation for gravity-accelerated atom-laser beams.
We can then take advantage of methods developped in optics and use the paraxial atom-optical $ABCD$ matrices formalism \cite{Borde1990,LeCoq2001}, instead of the general Kirchhoff integral, and characterize globally the beam with the quality factor $M^2$ \cite{Siegman1993,Belanger1991}.
\newline
\paragraph{The paraxial equation:}

We look for paraxial solutions to the time-independent
Schr\"{o}dinger equation,
\begin{equation}
    H_\mathbf{r}\psi_{\ell}(\mathbf{r})=E\psi_{\ell}(\mathbf{r})\,.
    \label{eq:SchrindepT}
\end{equation}

We decompose the wave function and the potential in a transverse
(``$\perp$") and parallel (``$\!/\!/$") component, taking $z$ as
the propagation axis,
\begin{equation}
\left\{
\begin{aligned}
\psi_{\ell}(x,y,z)&=\psi_{\perp}(x,y,z)\,\psi_{\!/\!/}(z)\,,\\
V(x,y,z)&=V_{\perp}(x,y,z)+V_{/\!/}(z)\,,
\end{aligned}
\right.
\label{eq:perpaarVpsi}
\end{equation}
where $V_{/\!/}(z)=V(0,0,z)$. We express the solution
$\psi_{\!/\!/}$ to the one dimensional equation
\begin{equation}
-\frac{\hbar^2}{2m}\frac{\partial^2 \psi_{\!/\!/}}{\partial z^2} +V_{/\!/}\psi_{\!/\!/}=E\psi_{\!/\!/}\,,
\label{eq:Schro1D}
\end{equation}
by using the WKB approximation,
\begin{equation}
\psi_{\!/\!/}(z)=\sqrt{\frac{m {\mathcal{F}} }{p(z)}}\exp\!\left[\frac{i}{\hbar}\int_{z_0}^z \mathrm{d}u\,p(u) \right]\,.
\label{eq:psiparr}
\end{equation}
In this expression, $\mathcal{F}$ is the atomic flux through any transverse plane, $p(z)=\sqrt{2m\left(E - V_{/\!/}(z)\right)}$ is the classical momentum along $z$ and $z_0$ is the associated classical turning point verifying $p(z_0)=0$. Using these expressions, and assuming an envelope $\psi_{\perp}$ slowly varying along $z$, we obtain the paraxial equation of propagation for the transverse profile,
\begin{equation}
    \left[i\hbar\partial_\zeta + \frac{\hbar^2}{2m}(\partial_x^2+\partial_y^2) - V_\perp(x,y,\zeta)\right] \psi_{\perp}(x,y,\zeta)=0\,,
    \label{eq:eqparaxial}
\end{equation}
where $\zeta(z) = \int_{z_0}^{z}\!dz\,m/p(z)$ is a parameter corresponding to the time which would be needed classicaly to propagate on axis from the turning point $z_0$.
The equation  \eqref{eq:eqparaxial} can thus be solved as a time-dependent Schr\"{o}dinger equation,
\begin{equation}
\psi_{\perp}(x,y,\zeta) =  \int_{\mathcal{S'}}\!\!dx' dy'\mathcal{K}(x,y;x'\!,y'\!;\zeta-\zeta')\,\psi_\perp(x'\!,y'\!,\zeta')\,.
\label{eq:eqparaxial2}
\end{equation}
The use of the paraxial approximation allows us to focus only on
the evolution of the transverse wave function, reducing the
dimensionality of the system from 3D to 2D, as the third dimension
along the propagation axis $z$ is treated via a semi-classical
approximation (Eq. \ref{eq:psiparr}).
\newline

\paragraph{$ABCD$ matrices:}

In the case of a separable transverse potential independent of
$z$, the paraxial approximation restricts to two independent one
dimensional equations. Let us consider a potential $V_x$ at most
quadratic in $x$. One can then write the propagator
$\mathcal{K}_x$ by using the Van Vleck formula, or equivalently
the general $ABCD$ matrix formalism \cite{Borde2001},
\begin{equation}
\mathcal{K}_x=\sqrt{\frac{\alpha}{2\pi i B}}\exp\left[\frac{i\alpha}{2B}\left(Ax'^2+Dx^2-2xx'\right)\right]\,.
\end{equation}
The coefficients $A$, $B$, $C$, $D$ verifying $AD-BC=1$, are functions of $\zeta-\zeta'$, and $\alpha$ is an arbitrary factor depending on the definition of the $ABCD$ coefficients. These ones are involved in the matrix describing the classical dynamics of a virtual particle of coordinate $X$ and speed $V$ in the potential $V_x(X)$
\begin{equation}
\begin{pmatrix}
 X(\zeta)\\ \alpha V(\zeta)
\end{pmatrix}=
\begin{pmatrix}
A(\zeta-\zeta') & B(\zeta-\zeta')/\alpha \\  \alpha C(\zeta-\zeta') & D(\zeta-\zeta')
\end{pmatrix}
\begin{pmatrix}
 X(\zeta')\\ \alpha V(\zeta')
\end{pmatrix}\,.
\label{eq:ABCDmat0}
\end{equation}
Different choices of $\alpha$ can be made and popular values in
the atomoptic literature are $\alpha=1$ \cite{Borde2001} or
$\alpha=m/\hbar$ \cite{LeCoq2001,Riou2006}. We take the last
convention and, by introducing the wavevector $K=mV/\hbar$, use
throughout this paper the following definition for the $ABCD$
coefficients in which is included the value of $\alpha$,
\begin{equation}
\begin{pmatrix}
 X(\zeta)\\ K(\zeta)
\end{pmatrix}=
\begin{pmatrix}
A(\zeta-\zeta') & B(\zeta-\zeta') \\ C(\zeta-\zeta') & D(\zeta-\zeta')
\end{pmatrix}
\begin{pmatrix}
 X(\zeta')\\ K(\zeta')
\end{pmatrix}\,.
\label{eq:ABCDmat}
\end{equation}
\newline

\paragraph{Propagation using the Hermite-Gauss basis:}
To calculate the propagation along the $x$ axis
\begin{equation}
\psi_x(x,\zeta)= \int\!\!dx' \mathcal{K}_x(x;x'\!;\zeta-\zeta')\,\psi_x(x',\!\zeta')\,,
\label{eq:propx}
\end{equation}
it is useful to use the Hermite-Gauss basis of functions $\left(\Phi_n\right)_{n \in \mathbb{N}}$
\begin{eqnarray}
    \Phi_0(x,\{X,K\}) =&\frac{(2\pi)^{-1/4}}{\sqrt{X}}\exp \left(i \frac{K}{X}\frac{x^2}{2}\right)\,, \\
    \Phi_n(x,\{X,K\}) =&\Phi_0(x) \frac{1}{\sqrt{2^n n!}} \frac{\left|X\right|^n}{X^n}H_n\left[\frac{x}{\sqrt{2}\left|X\right|}\right]\,.
    \label{eq:HGbasis}
\end{eqnarray}
$H_n$ is the $n$th order Hermite polynomial, and the two
parameters $(X,K)\in\mathbb{C}$, which define univocally the basis
set, must verify the normalization condition
\begin{equation}
KX^\ast-K^\ast X= i\,,
\label{eq:normcond}
\end{equation}
so that this basis is orthonormalized.

These functions propagate easily via $\mathcal{K}_x$, as
\begin{multline}
\Phi_n(x,\{X,K\}(\zeta))= \\ \int\!\!dx' \mathcal{K}_x(x;x'\!;\zeta-\zeta')\,\Phi_n(x'\!,\{X,K\}(\zeta'))\,,
\label{eq:propxPhi}
\end{multline}
\textit{i.e.} the integral is calculated by replacing $X(\zeta')$ and
$K(\zeta')$ by their value at $\zeta$ through the algebraic
relation \eqref{eq:ABCDmat}.

Thus, the propagation of the function $\psi_x$ between two positions $z(\zeta')$ and $z(\zeta)$ is obtained by first decomposing the initial profile on the Hermite-Gauss basis
\begin{equation}
\psi_x(x,\zeta')= \sum_n c_n \Phi_n(x,\{X,K\}(\zeta'))\,,
\label{eq:decompo1}
\end{equation}
where
\begin{equation}
c_n=\int dx \,\Phi_n^*(x,\{X,K\}(\zeta')) \,\psi_x(x,\zeta')\,.
\end{equation}
The profile after propagation until $z(\zeta)$ is then
\begin{equation}
\psi_x(x,\zeta)= \sum_n c_n \Phi_n(x,\{X,K\}(\zeta))\,.
\label{eq:recompo1}
\end{equation}
The high efficiency of this method comes from the fact that, once
the decomposition (\ref{eq:decompo1}) is made, the profile at
any position $z(\zeta)$ is obtained by calculating an algebraic
evolution equation: the $ABCD$ law (Eq. \ref{eq:ABCDmat}). Such
computational method is then much faster than the use of the
Kirchhoff integral, which would need to calculate an integral for
each considered position.

Note that the initial choice of $\{X,K\}(\zeta')$ is a priori
arbitrary as soon as it verifies the normalization condition (\ref{eq:normcond}). However, one can minimize the number of
functions $\Phi_n$ needed for the decomposition if one chooses
$\{X,K\}(\zeta')$ as a function of the second-order moments of the
profile.
\newline

\paragraph{Moments and quality factor:}

Let us define the second-order moments of $\psi_x$,
\begin{eqnarray}
\langle x x^\ast \rangle & = & \int dx \,x^2 \psi_x\psi_x^\ast\,, \label{eq:moments} \\
\langle k k^\ast \rangle & = & \int dx\,\partial_x \psi_x \,\partial_x \psi_x^\ast\,, \\
\langle x k^\ast + x^\ast k \rangle & = & i \int dx\, x\left[\psi_x\partial_x \psi_x^\ast - \psi_x^\ast\partial_x\psi_x\right] \,,
\end{eqnarray}
where we have used that $\psi_x$ is normalized \mbox{($\int dx |\psi_x|^2=1$).} We also define the wavefront curvature $\mathcal{C}$ \cite{Siegman1991} as
\begin{equation}
    \mathcal{C} = \frac{ \langle x k^\ast + x^\ast k \rangle}{2 \langle x x^\ast \rangle}\,.
    \label{eq:curvature}
\end{equation}

The three moments follow also an $ABCD$ law during propagation. By introducing the matrix
\begin{equation}
\mathcal{M}(\zeta) =
\begin{pmatrix}
        \langle x x^\ast\rangle                     & \langle x k^\ast + x^\ast k \rangle/2 \\
        \langle x k^\ast + x^\ast k \rangle/2   & \langle k k^\ast\rangle
\end{pmatrix}\,,
\end{equation}
this law is expressed as
\begin{equation}
    \mathcal{M}(\zeta)=
    \begin{pmatrix}
        A & B \\  C & D
    \end{pmatrix}
    \mathcal{M}(\zeta')
    \begin{pmatrix}
        A & B \\  C & D
    \end{pmatrix}^t\,.
\end{equation}

This relation allows to derive propagation laws on the wavefront second-order moments, such as the r.m.s
transverse size (Rayleigh law). As $\mathrm{det} (\mathcal{M})$ is constant, this law also exhibits an invariant
of propagation, the beam quality factor $M^2$, related to the moments and curvature by
\begin{equation}
 \langle x x^\ast \rangle \left( \langle k k^\ast \rangle - \mathcal{C}^2 \langle x x^\ast \rangle \right)=\left(\frac{M^2}{2}\right)^2 \,.
\label{eq:M2gene}
\end {equation}
The physical meaning of the $M^2$ factor becomes clear by taking the last equation at the waist, \textit{i.e.} where the
curvature $\mathcal{C}$ is zero:
\begin{equation}
    \sqrt{\langle x x^\ast \rangle_0 \langle k k^\ast \rangle_0} = \frac{M^2}{2}\,.
    \label{eq:M2}
\end {equation}
The $M^2$ factor is given by the product of the spatial and momentum widths at the beam waist and indicates
how far the beam is from the diffraction limit. Because of the Heisenberg uncertainty relation, the $M^2$
factor is always larger than one and equals unity only for a perfect gaussian wavefront.

Finally, the determination of the second order moments and the $M^2$ factor from an initial profile allows us to
choose the more appropriate values of $\{X(\zeta'),K(\zeta')\}$ to parameterize the Hermite-Gauss basis used for
the decomposition at $z(\zeta')$ (Eq. \ref {eq:decompo1}). Indeed, these parameters are closely related to the
second order moments of the Hermite-Gauss functions $\Phi_n(x, \{X,K\})$ by
\begin{eqnarray}
\langle x x^\ast \rangle_{\Phi_n} & = & (2n+1)\left|X\right|^2\,, \label{eq:X}\\
\langle k k^\ast \rangle_{\Phi_n} & = & (2n+1)\left|K\right|^2\,, \\
\langle xk^\ast + x^\ast k \rangle_{\Phi_n} & = & (2n+1) \left( X K^\ast + X^\ast K \right)\,.
\end{eqnarray}
From this we obtain that the $M^2$ factor of the mode $\Phi_n$ is \mbox{$M^2_{\Phi_n} = (2n+1)$} and that all the modes have the same curvature
\begin{equation}
\mathcal{C}_{\Phi_n}=\mathcal{C}=\left( X K^\ast + X^\ast K \right)/2|X^2|\,. \label{cond:curv}
\end{equation}
It is thus natural to choose the parameters $\{X,K\}$, so that the curvature of the profile (Eq.
\ref{eq:curvature}) equals $\mathcal{C}$. This last condition, together with the choice $|X|^2=\langle x x^\ast \rangle/M^2$, the normalization condition (\ref{eq:normcond}) and
the choice of $X$ real (the phase of $X$ is a global phase over the wavefront), lead to the univocal determination of the parameters $\{X(\zeta'),K(\zeta')\}$ associated with the Hermite-Gauss basis, so that the decomposition of the
initial profile $\psi_x(x,\zeta')$ needs a number of terms of the order of $M^2$.

\section{Application to a radiofrequency-outcoupled atom laser}

We apply the previous framework to the radiofrequency (rf) outcoupled atom
laser described in \cite{Riou2006} where a Bose-Einstein condensate
(BEC) of rubidium 87 (mass $m$) is magnetically harmonically
trapped (frequencies $\omega_x=\omega_z=\omega_\perp$ and
$\omega_y$) in the ground state $\left|F=1,m_F=-1\right\rangle$,
and is weakly outcoupled to the untrapped state
$\left|F=1,m_F=0\right\rangle$. The BEC
is considered in the Thomas-Fermi (TF) regime described by the time-independent wave function $\phi_\mathrm{s}(r)$, with a
chemical potential $\mu$ and TF radii $R_{\perp,y}=\sqrt{2\mu /
m\omega^2_{\perp,y}}$ \cite{Dalfovo1999}. The external potential experienced by the
beam is written
\begin{equation}\label{eq:Vi}
V_\mathrm{i}(r) = \mu
-\frac{1}{2}m\omega_\perp^2\sigma^2-\frac{1}{2}m
\left[\omega_\perp^2(x^2+z^2)+\omega_y^2 y^2\right]
\end{equation}
inside the BEC region, and
\begin{equation}\label{eq:Vo}
V_\mathrm{o}(r) = \frac{1}{2} m\omega^2\sigma_q^2 -
\frac{1}{2}m\omega^2 \left[x^2+(z+\sigma_q)^2\right].
\end{equation}
outside. The expulsive quadratic potential of $V_\mathrm{i}$
originates from the mean-field interaction (independent of the
Zeeman substates for $^{87}$Rb) between the laser and the
condensate, whereas that of $V_\mathrm{o}$ (frequency $\omega$) is
due to the second order Zeeman effect. We have noted $\sigma =
g/\omega_\perp^2$ and $\sigma_q=g/\omega^2$ the vertical sags due
to gravity $-mgz$ for $m_F=-1$ and $m_F=0$ states respectively.
The rf coupling (of Rabi frequency $\Omega_R$) between the
condensate and the beam is considered to have a negligible
momentum transfer and provides the atom-laser wave function with a
source term $\rho = \hbar\Omega_R/2 \,\phi_\mathrm{s}(r)$.

\begin{figure}[bt]
\centering
\scalebox{1}{\includegraphics{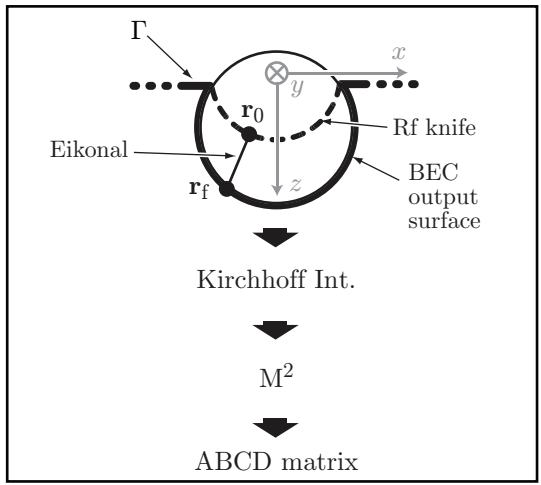}}
\caption{Principle of the calculation : the wave function is calculated from the
rf knife (a circle of radius $r_0$ centered at the frame origin) using the eikonal. A general radial atomic trajectory starting at zero
speed from $\mathbf{r}_0$  crosses the BEC border at $\mathbf{r}_{\mathrm{f}}$. Once the matter wave has exited the condensate region, the wave function is
given by the Fresnel-Kirchhoff integral,
allowing to compute the wave function at any point 
from the BEC output surface. In the paraxial approximation,
we calculate the propagation using  $ABCD$ matrices.
}
\label{principlecalc}
\end{figure}

In the following, we consider a condensate elongated along
the $y$ axis ($\omega_\perp \gg \omega_y$), so that the laser
dynamics is negligible along this direction \footnote{This has
been checked in a similar configuration by numerical simulations
as reported in \cite{Kohl2005}.}. We thus study independently the
evolution in each vertical $(x,z)$ plane at position $y_0$. We calculate the
beam wave function in two steps corresponding to a
propagation in each region defined by $V_\mathrm{i}$ and
$V_\mathrm{o}$ (see Fig. \ref{principlecalc}). The wave function at
the BEC frontier is calculated in section \ref{propaginBEC} using
the eikonal approximation. Then, in section \ref{KirchI}, we
calculate the wave function at any position outside the BEC, with
the help of the Fresnel-Kirchhoff formalism and the paraxial ABCD
matrices.

\subsection{Propagation in the condensate zone}
\label{propaginBEC}

In this section, we determine the beam wave function
$\psil(\mathbf{r})$ in the condensate zone by using the eikonal
formalism described in section \ref{sec:eikonal}. This formalism is appropriate in this case
as the time necessary for the laser to exit the
BEC region ($\approx$ 1 ms) is small enough so that the transverse diffraction is negligible 
(transverse size $\approx R_{\perp}$).

\subsubsection{Atomic rays inside the BEC}

One first needs to calculate the atomic paths followed by the atom
laser rays from the outcoupling surface (the rf knife) to the border of the BEC. 
The rf knife is an ellipso\"{i}d centered at the
magnetic field minimum (chosen in the following as the frame origin, see figure
\ref{principlecalc}). Its intersection with the $(x,z)$ plane at position $y_0$ 
is a circle centered at the frame origin. Its radius $r_0$ depends on the rf
detuning $\delta\nu = \frac{m}{2h}\left[\omp^2 \left({r_0}^{\!2}
-\sigma^2\right)+\omega_y^2 {y_0}^{\!2}\right]$.
As we neglect axial dynamics and consider zero initial momentum, 
the classical equations of motion give for the radial coordinate $r=\sqrt{x^2+z^2}$,
$r(t)=r_0\cosh{\omega_{\perp}t}$ allowing  to find a starting
point $\mathbf{r}_0$  on the rf knife for each point $\mathbf{r}_{\mathrm{f}}$
on the BEC output surface, \textit{i.e.} the BEC border below the rf knife
\cite{PhDThesisRiou}.

\subsubsection{Eikonal expression of the wave function}

We now introduce
\begin{equation}
a_{\perp}=\sqrt{\frac{\hbar}{m\omega_{\perp}}},\,\,
R=\frac{r}{a_{\perp}},\,\,\epsilon=-\left(\frac{r_0}{a_{\perp}}\right)^{\!2}\!\!,
\label{eqparam}
\end{equation}
which are respectively the size of the harmonic potential, the
dimensionless coordinate and energy  associated with the atom
laser. Following Eq. \eqref{eq:Seikonal} and \eqref{eq:Aeikonal},
we obtain
\begin{equation}
\left\{
\begin{aligned}
S(R) & = \frac{\hbar}{2} \left[ R\sqrt{R^2+\epsilon}+\epsilon \ln \left(\frac{R+\sqrt{R^2+\epsilon}}{\sqrt{-\epsilon}}\right)\right],\\
A(R)& = \frac{{\cal{B}}
\phi_{\mathrm{s}}(\mathbf{r}'_0)}{\left[R^2(R^2+\epsilon)\right]^{1/4}}.
\end{aligned}
\right. \label{WKB}
\end{equation}
${\cal{B}}$ is proportional to the
coupling strength, and is not directly given by the eikonal
treatment \footnote{A detailed development of the first term of Eq.
\ref{eq:MagicRuleIndepT} using the stationary phase approximation
leads to the value ${\cal{B}}\!=\!\Omega_{\mathrm{R}}/\omp \sqrt{i\pi/2}$
\cite{PhDThesisRiou}. However, in practice, as long as one is
interested only in the shape of the wave function, the precise
value of $C$ is not necessary.}.
The atom laser beam amplitude $A(R)$ \footnote{
This treatment does not allow us to predict the wave function
value in the vicinity of the classical radial turning-point
$R_0=\sqrt{-\epsilon}$ since the normalisation $A(R)$ diverges as
$R\rightarrow R_0$. A more accurate treatment is thus needed to
deal with the few trajectories starting at the edge of the rf
knife. One possibility is to use the exact solution $\psi$ of the
radially symmetric two-dimensional inverted potential, as
presented in appendix \ref{invoh}.}
is proportional to the BEC wave function value at the rf knife
$\phi_{\mathrm{s}}(\mathbf{r}_0)$.
The wave function at the BEC output surface is then
\begin{equation}
\psi_{\ell}(R_{\mathrm{f}})=
A(R_{\mathrm{f}})\exp\left[\,iS(R_{\mathrm{f}})/\hbar\,\right].
\label{eqpartI}
\end{equation}

\subsection{Propagation outside the condensate}
\label{KirchI}

Once the matter wave has exited the condensate region, the volume
source term $\rho$ vanishes and the beam wave function is
given by the second term of Eq. \ref{eq:MagicRuleIndepT} only,
\textit{i.e.} the Fresnel-Kirchhoff integral for matter waves,
allowing to compute the wave function at any point from the wave
function on the BEC output surface. In this section, we calculate the propagation using an analytical expression for the
time-independent propagator and apply the $ABCD$ formalism in the paraxial regime.

\subsubsection{Fresnel-Kirchhoff Integral}

We
perform the Fresnel-Kirchhoff integral in the $(x,z)$ plane at position $y_0$:
\begin{equation}
\psi_{\ell}=\frac{i\hbar}{2m} \oint_{\Gamma} {d\bf{l}'} \cdot
\bigl[{\cal{G}}_E \nabla \psi_{\ell} -\psi_{\ell} \nabla
{\cal{G}}_E \bigr], \label{Kirchhoff}
\end{equation}
where $\psi_\ell$ is non zero only on the
BEC output surface as seen in section \ref{propaginBEC}.
The surface $\mathcal{S}$ of equation \eqref{eq:MagicRuleIndepT}
is here reduced to its interserction contour $\Gamma$ with the vertical plane.
It englobes the BEC volume and is closed at
infinity.

\begin{figure}[bt]
\centering \scalebox{1}{\includegraphics{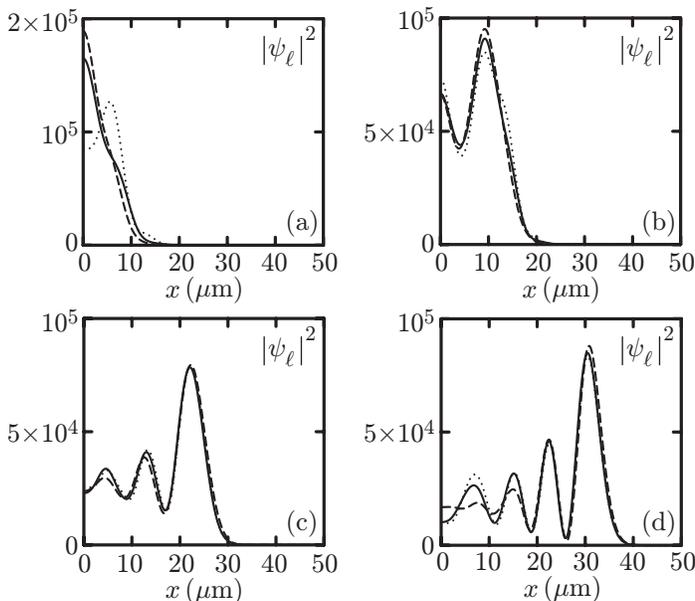}}
\caption{Density profiles obtained at $150~\mu\mathrm{m}=z-\sigma$
below the BEC center. We consider the vertical plane $y_0=0$ and
have normalised $|\psil|^2$ to unity. We have drawn the results
obtained by using as input of the Kirchhoff integral the profile
calculated using the eikonal (Eq. \ref{eqpartI}, dotted line) or
exact solutions of the inverted harmonic oscillator (Eq.
\ref{ExprWhitt}, full line), and compare them to a full numerical
integration of the two-dimensional Gross-Pitaevskii evolution of
the atom laser (dashed line). The used rf detunings are: (a)
$\delta \nu=~ 8900$ Hz, (b) $\delta \nu=~ 6500$ Hz, (c) $\delta
\nu=~ 2100$ Hz, (d) $\delta \nu=-1100$ Hz, and correspond to
increasing outcoupling height, from (a) to (d).} \label{FIG4}
\end{figure}

Using the expression of 
${\cal G}_E$  calculated in appendix \ref{sec:PropagatorindepQuadratic}, 
we compute equation (\ref{Kirchhoff}) and the result is shown in Fig. \ref{FIG4} for four different
outcoupling rf detunings. When coupling occurs at the top of the
BEC, the propagation of the beam exhibits a strong
divergence together with a well-contrasted interference pattern.
The divergence is due to the strong expulsive potential
experienced by the beam when crossing the condensate, and
interferences occur because atomic waves from different initial
source points overlap during the propagation.

Comparison with a  numerical Gross-Pitaevskii simulation shows
good agreement. We also compare the results obtained by using at
the BEC surface either Eq. \ref{eqpartI} or Eq. \ref{ExprWhitt}.
The eikonal method fails when coupling at the very bottom of the
BEC [Fig. \ref{FIG4}(a)], since the classical turning point is too
close to the BEC border, whereas the method using the exact
solutions of the inverted harmonic potential agrees much better
with the numerical simulation for any rf detuning. Finally, for
very high coupling in the BEC [Fig. \ref{FIG4}(d)], our model
slightly overestimates the fringe contrast near the axis.

\subsubsection{Propagation in the paraxial regime}
\label{ABCDm}

Since the atom laser beam is accelerated by gravity, it enters quickly
the paraxial regime. In the case considered in \cite{Riou2006}, the
maximum transverse energy is given by the chemical potential $\mu$
whereas the longitudinal energy is mainly related to the fall
height $z$ by $E_z \approx mgz$. For $\mu$ typically of a few kHz,
one enters the paraxial regime after approximately $100~\mu$m of
vertical propagation. For larger propagation distances, we can
thus take advantage of the paraxial approximation presented in
Sec. \ref{propagparax}.

To proceed, we start from the profile $\psi_{\ell}(x)$ calculated
after 150 $\mu \mathrm{m}$ of propagation via the Kirchhoff
integral. Using Eqs. \eqref{eq:moments}-\eqref{eq:curvature}, we
extract the widths $<xx^*>$, $<kk^*>$ and the beam curvature $\cal{C}$
at this position. From these parameters we calculate
the beam quality factor $M^2$ by using the general equation
\eqref{eq:M2gene}.
Following the procedure presented in Sec. \ref{propagparax}, we
can choose the appropriate Hermite-Gauss decomposition of $\psi_{\ell}(x)$ and
the propagation of each mode is then deduced
from the $ABCD$ matrix corresponding to the transverse part of the potential
described in 
Eq. \ref{eq:Vo}, $ V_{\perp}(x)=-\frac{m}{2}\omega^2x^2$.
The $ABCD$ matrix then reads
\begin{equation}
\begin{pmatrix}
A & B \\  C & D
\end{pmatrix}
 = \begin{pmatrix}\cosh{\omega (\zeta-\zeta')}& \frac{\hbar}{m\omega}\sinh{\omega(\zeta-\zeta')} \\ \frac{m\omega}{\hbar}\sinh{\omega (\zeta-\zeta')}& \cosh{\omega
 (\zeta-\zeta')}
\end{pmatrix}\, .
\end{equation}
As explained in Sec. \ref{propagparax}, the propagation is
parameterized by the time $\zeta$, given by the classical equation
of motion of the on-axis trajectory in the longitudinal part of the potential
$V_{\parallel }({\tilde z})=-\frac{m}{2}\omega^2{\tilde
z}^2$,
where $\tilde z= z+\sigma_q$. 

The $ABCD$ matrices formalism allows also to extract global
propagation laws on the second order moments $X(\zeta)$,
$K(\zeta)$ and evaluate the wavefront curvature
$\mathcal{C}(\zeta)=\Re \left(\frac{K(\zeta)}{X(\zeta)}\right)$
associated with the wavefront $\psi_{\ell}(x,\zeta)$.

By considering the paraxial evolution of the r.m.s.
size $\sigma$ of  $\psi_{\ell}(x,\zeta)$, we then obtain 
a generalized Rayleigh formula :
\begin{equation}
\sigma^2(\xi)=\sigma_{0}^2\cosh^2\left(\omega\xi\right)+\left(\frac{M^2\hbar}{2m\omega}\right)^2
\frac{\sinh^2\left(\omega\xi\right)}{\sigma_{0}^2}\,,
\label{sigma}
\end{equation}
involving the $M^2$ factor, and where
$\sigma_0 = X(\zeta_0)$ and $\xi = \zeta - \zeta_0$. 
We have introduced the focus time $\zeta_0$ so that $\mathcal{C}(\zeta_0)=0$.
The relation \eqref{sigma} has been
fruitfully used in \cite{Riou2006} and \cite{Jeppesen2007} to extract the beam quality
factor from experimental images.

\section*{Conclusion}

Relying on the deep analogy between light waves and matter waves,
we have introduced theoretical tools to deal with the
propagation of coherent matter waves :
\begin{itemize}
\item The eikonal approximation is the standard treatment of
geometrical optics. It is valid when diffraction, or wave-packet
spreading, is negligible. It can be fruitfully used to treat short
time propagation, as we show on the exemple of an atom laser beam
crossing its source BEC. 
\item The Fresnel-Kirchhoff integral comes from
the classical theory of diffraction. It is particularly powerful
as it allows to deal with piecewise defined potential in two or
three dimensions together with taking into account diffraction and
interference effects. 
\item The $ABCD$ matrices formalism can be used as soon as the matter wave
is in the paraxial regime.
This widely used technique in laser optics provides simple
algebraic laws to propagate the atomic wavefront, and also global
laws on the second order moments of the beam, as the Rayleigh
formula. Those results are especially suitable to characterize
atom laser beams quality by the $M^2$ factor. 
\end{itemize}
 The toolbox developed in
this paper can efficiently address a diversity of atom-optical
setups in the limit where interactions in the laser remain negligible. 
It can be suited for beam focussing experiments 
\cite{Shvarchuck2002,Arnold2004} and their
potential application to atom lithography
\cite{Knyazchyan2005}.
It also provides a relevant insight on beam profile effects in
interference experiments involving atom lasers or to characterize
the outcoupling of a matter-wave
cavity \cite{Impens2006}. It could also be used in estimating
the coupling between an atom laser beam and
a high finesse optical cavity \cite{Ritter2007}.
Further
developments may be carried out to generalize our work. In particular, 
the $M^2$ factor approach could be generalized to self interacting atom laser beam in
the spirit of \cite{Pare1992} or to more general cases of
applications, such as non-paraxial beams or more complex external
potential symmetries \cite{Impens2008}.\\

\begin{acknowledgments}

The LCFIO and SYRTE are members of the Institut Francilien de
Recherche sur les Atomes Froids (IFRAF). This work is supported by
CNES (No. DA:10030054), DGA (Contracts No. 9934050 and No.
0434042), LNE, EU (grants No. IST-2001-38863, No.
MRTN-CT-2003-505032 and FINAQS STREP), ESF (No. BEC2000+ and
QUDEDIS).

\end{acknowledgments}

\appendix

\section{Time-independent propagator in an inverted harmonic potential}
\label{sec:PropagatorindepQuadratic}

The time-dependent propagator of the inverted harmonic potential can be straightforwardly deduced from its expression for the harmonic potential \cite{Feynman1965} by changing real trapping frequencies to imaginary ones ($\omega\rightarrow i \omega$). We derive here an analytic evaluation of its time-independent counterpart ${\cal G}_{E}$, by using the results of section
\ref{GreenGeneral}.

We consider a potential in dimension $d$, characterized by the expulsing frequency $\omega$
\begin{equation}
V(\mathbf{r})=V(\mathbf{0})-\!\sum_{j\in
[\![1..d]\!]}\frac{1}{2}m\omega^2r_j^2.
\end{equation}
By introducing the reduced time $s=\omega \tau$ and the harmonic oscillator size $\sigma_{\mathrm{o}}=\sqrt{\hbar/m\omega}$, ${\cal G}_E$ is expressed as
\begin{equation}
{\cal G}_E( {\mathbf{r}},{\mathbf{r}}')
=\int_{0}^{\infty} ds \, {\cal H}(s)
e^{i\phi({\mathbf{r}},{\mathbf{r}}',s)}, \label{GE}
\end{equation}
with ${\cal H}(s)=m/(2\pi i \hbar \sinh s)$,
and
\begin{equation}
\phi= \frac{ \left[\bigl({\mathbf{r}}^2 +
{\mathbf{r}}'^2\bigr)\cosh s
-2{\mathbf{r}}\cdot{\mathbf{r}}' \right]}{2
\sigma_{\mathrm{o}}^2 \sinh s}
+\frac{\left(E-{V}(\mathbf{0})\right)}{\hbar\omega}\,s.
\end{equation}
The first-order stationary times $s_{\pm}$ verify
\begin{equation}
\cosh{s_\pm}= \frac{-b \pm
\sqrt{b^2+4(E-{V}(\mathbf{0}))c}}{2(E-{V}(\mathbf{0}))},
\label{stat1}
\end{equation}
where $b=m \omega^2{\mathbf{r}}\cdot{\mathbf{r}}'$ and $c=E-{V}(\mathbf{0})+m\omega^2
({\mathbf{r}}^2 + {\mathbf{r}}'^2)/2$.
If there are positive and real solutions $s_\pm$, ${\cal G}_E$
reads (eq. \ref{eq:GE1})
\begin{equation}
{\cal G}_E^{(1)}({\mathbf{r}},{\mathbf{r}}') =
\sum_{s_\pm>0}  \sqrt{\frac{2i\pi}{\partial^2 \phi/\partial
s^2\bigr|_{s_\pm}}} {\cal H}(s_\pm ) e^{i\phi(s_\pm)}. \label{GE1}
\end{equation}

Otherwise, the relevant stationary point $s_0$ (eq. \ref{eq:defstat2}) verifies 
\begin{equation}
\cosh s_0=\frac{{\mathbf{r}}^2+{\mathbf{r}}'^2+\sqrt{
\left({\mathbf{r}}+{\mathbf{r}}'\right)^2\left({\mathbf{r}}-{\mathbf{r}}'\right)^2}
}{2\, {\mathbf{r}} \cdot {\mathbf{r}}'}. \label{s0}
\end{equation}
$s_0$ is the time associated with the classical trajectory
connecting ${\mathbf{r}}'$ and ${\mathbf{r}}$ with the
closest energy to $E$. If the angle beetween ${\mathbf{r}}$
and ${\mathbf{r}}'$ is above $\pi/2$, then according to
equation (\ref{s0}), the absolute value of the first derivative of
$\phi$ is never minimal, so that $e^{i\phi(s)}$ quickly oscillates
over $[0;+\infty)$ and one can take ${\cal
G}_E({\mathbf{r}},{\mathbf{r}}')=0$. In other cases,
where the solution is unique, one develops the phase around $s_0$
and ${\cal G}_E$ finally expresses as (eq. \ref{eq:GE2})
\begin{equation}
{\cal G}_E^{(2)}({\mathbf{r}},{\mathbf{r}}')=\frac{2
\pi {\cal H}(s_0)}{ \kappa}e^{i\phi(s_0)}
\mathrm{Ai}\left(-\frac{1}{\kappa} \frac{\partial \phi}{\partial
s}\Bigr|_{s_0}\right), \label{GE2}
\end{equation}
where $\kappa=\left(-\frac{1}{2}\frac{\partial^3 \phi}{\partial
s^3}\bigr|_{s_0}\right)^{1/3}$.

\section{Exact solutions of the two-dimensional inverted harmonic oscillator and relation with the eikonal}
\label{invoh}

In this appendix, we give an analytical expression for the
eigenfunctions of the inverted harmonic potential in the BEC
region. The use of such solutions enable us to avoid any
divergence of the eikonal solution close to the turning point.

Using dimensionless parameters introduced in Eq. \eqref{eqparam},
the time-independent Schr\"{o}dinger equation in the BEC region
reads \begin{equation}
-\left(\frac{\partial^2\psi}{\partial
R^2}\!+\frac{1}{R}\frac{\partial\psi}{\partial
R}+\!\frac{1}{R^2}\frac{\partial^2\psi}{\partial
\alpha^2}\right)\!-\!R^2\psi=\epsilon \psi.
\end{equation}
Introducing the angular momentum
$L_{\alpha}=\frac{\hbar}{i}\frac{\partial \psi}{\partial \alpha}$,
one can decompose the solution of this equation as the product of
a radial part and an angular part
\begin{equation}
\psi(R,\alpha)=\phi(R)\, e^{i l \alpha},
\end{equation}
with $l \in  \mathbb{Z}$ and $\hbar l$ is the angular momentum of
the wave function. The general solution $\phi$ is given by
\begin{equation}
\phi(R)=\frac{c_1}{R}\mathrm{M}\!\left(-i\frac{\epsilon}{4};\frac{l}{2};iR^2
\right)\!+\frac{c_2}{R}\mathrm{W}\!\left(-i\frac{\epsilon}{4};\frac{l}{2};iR^2
\right)\!. \label{whittaker}
\end{equation}
$\mathrm{M}(\mu,\nu, z)$ and $\mathrm{W}(\mu,\nu, z)$ are
Whittaker functions (related to the confluent hypergeometric
functions of the first and second kind) \cite{Abramowitz1972}
whereas $c_1$ and $c_2$ are complex coefficients.

In general, the wave function must be decomposed on the basis of
the different solutions $\phi(R)$ parameterized by $l$ and
$\epsilon$. However, in the following, we restrict ourselves to
the study of a solution that connects asymptotically to the
eikonal. Thus, we are only interested in the wave function
describing a dynamics without any transverse speed or diffraction,
\textit{i.e.} with $l=0$. Since the wave progresses from the rf knife
$R_0$ to the outer part of the potential, we also only look for
``outgoing wave" type solutions \cite{Fertig1987}. Such solutions
behave as progressive waves in the asymptotic limit ($R\rightarrow
\infty$). One can express the Whittaker functions in term of
hypergeometric functions \cite{Abramowitz1972} for any complex
parameter $\mu$ and $z$
\begin{eqnarray}
\mathrm{M}(\mu,0, z)&=&e^{-z/2}\sqrt{z}{}_1\mathrm{F}^1\!\left(\frac{1}{2}-\mu;1;z\right)\!, \\
\mathrm{W}(\mu,0,z)&=&e^{-z/2}z^{\mu}{}_2\mathrm{F}^0\!\left(\frac{1}{2}-\mu,\frac{1}{2}-\mu;;-\frac{1}{z}\right)\!.
\end{eqnarray}
For $|z| \rightarrow \infty$, these functions are asymptotically
expanded as \cite{Landau1977}
\begin{multline}
{}_1\mathrm{F}^1(a;b;z)\sim \frac{\Gamma(b)}{\Gamma(b-a)}(-z)^{-a}{}_2\mathrm{F}^0\left(a,a-b+1;;-\frac{1}{z}\right)\\
+\frac{\Gamma(b)}{\Gamma(a)}e^z
z^{a-b}{}_2\mathrm{F}^0\left(b-a,1-a;;\frac{1}{z}\right)\!,
\end{multline}
and
\begin{equation}
{}_2\mathrm{F}^0\left(a,b;;\frac{1}{z}\right)\longrightarrow
1+{\cal O}\left(\frac{1}{z}\right)\!.
\end{equation}

One thus obtains an asymptotic formula for equation
(\ref{whittaker}) in which terms proportional to $e^{iR^2/2}$ or
$e^{-iR^2/2}$  appear. Cancelling the second ones corresponding to
an incoming wave towards the center, leads to a relation
between $c_1$  and $c_2$:
\begin{equation}
i\frac{e^{-\pi \epsilon/4}}{\Gamma(\frac{1}{2}-i\frac{\epsilon}{4}
)}c_1+c_2=0
\end{equation}
The solution is finally written as:
\begin{widetext}
\begin{equation}
\psi(R)=\frac{\Gamma\left(\frac{1}{2}+i\frac{\epsilon}{4}\right)e^{i\epsilon\left[1-\ln(-\epsilon/4)\right]/4}}{R}\biggl[e^{\pi\epsilon/8}\mathrm{M}\left(-i\frac{\epsilon}{4};0;iR^2
\right)-\frac{ie^{-\pi
\epsilon/8}}{\Gamma(\frac{1}{2}-i\frac{\epsilon}{4}
)}\mathrm{W}\left(-i\frac{\epsilon}{4};0;iR^2 \right)\biggr].
\label{ExprWhitt}
\end{equation}
\end{widetext}
where the prefactor has been chosen so that the asymptotic
expression of $\psi(R)$ connects to the eikonal solution given by
equation \eqref{WKB}.


\end{document}